\documentclass{elsart}

\usepackage{graphics}
\usepackage{epsf}
\usepackage{epsfig}
\usepackage{amssymb}

\journal{Nuclear Physics B}
\volume{718}
\firstpage{341}
\lastpage{361}
\pubyear{2005}

\received{2 March 2005}
\accepted{29 April 2005}

\begin{document} 

\begin{frontmatter} 


\title{Scaling behavior of the directed percolation universality class}

\author[LUEB]{{S. L\"ubeck}\corauthref{cor}},
\corauth[cor]{Corresponding author.}
\ead{sven@thp.uni-duisburg.de}
\author[WILL]{R.\,D. Willmann}
\ead{r.willmann@fz-juelich.de}

\address[LUEB]{Theoretische Physik, 
Universit\"at Duisburg-Essen,
47048 Duisburg, Germany 
}

\address[WILL]{Institut f\"ur Festk\"orperforschung,
Forschungszentrum J\"ulich,
52425 J\"ulich, Germany 
}



\begin{abstract}
In this work we consider five different lattice models
which exhibit continuous phase transitions into absorbing
states.
By measuring certain universal functions, which
characterize the steady state as well as the dynamical
scaling behavior, we present clear numerical evidence
that all models belong to the universality class of
directed percolation.
Since the considered models are characterized by different
interaction details the obtained universal scaling
plots are an impressive manifestation of the universality
of directed percolation.
\end{abstract}


\end{frontmatter}



\section{Introduction}

In this work we consider the universality class of directed
percolation (DP, see~\cite{HINRICHSEN_1,ODOR_1} for recent reviews).
Because of its robustness and ubiquity,
(including critical phenomena in physics, biology, epidemiology,
as well as catalytic chemical reactions) directed
percolation is recognized as the paradigm of non-equilibrium
phase transitions into absorbing states.
These so-called absorbing phase transitions arise from a 
competition of opposing processes, 
usually creation and annihilation processes.
The transition point separates an active phase
from an absorbing phase in which the dynamics is frozen.
Analogous to equilibrium critical phenomena, absorbing
phase transitions can be grouped into different universality
classes.
All systems belonging to a given universality class
share the same critical exponents, and certain
scaling functions (e.g.~equation of state, correlation 
functions, finite-size scaling functions, etc.) 
become identical near the critical point.
According to the 
universality hypothesis of Janssen and Grassberger, 
short-range interacting models, 
exhibiting a continuous phase transition into a unique 
absorbing state, belong to the directed percolation 
universality class, if 
they are characterized by a one-component order 
parameter and no additional symmetries~\cite{JANSSEN_1,GRASSBERGER_2}.

Similar to equilibrium critical phenomena, the universality
of directed percolation is understood by renormalization
group treatments of an associated continuous field theory.
The process of directed percolation might be 
represented by the Langevin
equation~\cite{JANSSEN_1} 
\begin{equation}
\partial_{\scriptscriptstyle t} \, n({\underline x},t)
\; = \; r \, n({\underline x},t) 
\, - \, u \, n^2({\underline x},t) 
\, + \, \Gamma \, \nabla^2 \, n({\underline x},t) 
\, + \, 
\eta({\underline x},t)  \, .
\label{eq:langevin_dp_01}
\end{equation}
Here, $n({\underline x},t)$ corresponds to the density of 
active sites on a mesoscopic scale and $r$~describes the
distance to the critical point.
Furthermore, $\eta$ denotes the 
noise which accounts for fluctuations of $n({\underline x},t)$.
According to the central limit theorem, 
$\eta({\underline x},t)$ 
is a Gaussian random variable with zero mean
and whose correlator is given by
\begin{equation}
\langle \, \eta({\underline x},t) \, 
\eta({\underline x}^{\prime},t^{\prime}) \, \rangle
\; = \; \kappa \; n({\underline x},t) \;
\delta({\underline x}-{\underline x}^{\prime}) \;
\delta(t-t^{\prime}) \, .
\label{eq:langevin_dp_corr_01}
\end{equation}
Notice, that the latter equation ensures that the 
system is trapped in the absorbing state 
$n({\underline x},t)=0$.
Furthermore, higher order terms such as 
$n({\underline x},t)^3$, $n({\underline x},t)^4$ 
or $\nabla^4n({\underline x},t)$
are irrelevant under renormalization group transformations
as long as $u > 0$.
Negative values of $u$ give rise to a first order
phase transition whereas $u=0$ is associated with
a tricritical point~\cite{OHTSUKI_1,OHTSUKI_2}.

Above the upper critical 
dimension~$D_{\scriptscriptstyle \mathrm c}$
mean field theories apply and present instructive insight
into the critical behavior~\cite{MORI_1}.
Below~$D_{\scriptscriptstyle \mathrm c}$, 
renormalization group techniques have to be applied
to determine the critical exponents and the
scaling functions 
(see~\cite{JANSSEN_13,TAUBER_2} for recent reviews of the 
field theoretical treatment of 
directed percolation).
In that case path integral formulations are more
adequate than the Langevin equation.
Stationary correlation functions as well as response
functions can be determined by calculating 
path integrals with weight $\exp{(-{\mathcal S})}$,
where the dynamic functional ${\mathcal S}$ describes
the considered stochastic process.
Up to higher irrelevant orders the dynamic functional 
associated with directed percolation is given 
by~\cite{JANSSEN_1,JANSSEN_7,DEDOMINICIS_1,JANSSEN_8} 
\begin{equation}
{\mathcal S}[{\tilde n},n] \; = \; \int
{\mathrm d}^D{\underline x} \,
{\mathrm d}t \;
{\tilde n}  \, \left [ \, 
\partial_{\scriptscriptstyle t} n  -  (r + \nabla^2 ) n
 -  \left (  \frac{\kappa}{2} \, {\tilde n}  - u \, n \right ) n \, 
\right ]
\label{eq:action_reggeon_field_theory}
\end{equation}
where ${\tilde n}({\underline x},t)$
denotes the response field conjugated to the 
Langevin noise field~\cite{MARTIN_1}.
Remarkably, the above functional is well known from
high energy physics and corresponds to the 
Lagrangian of Reggeon field theory~\cite{ABARBANEL_1}.
Since DP represents the simplest realization of a nonequilibrium
phase transition its field theory is often regarded as
the nonequilibrium counterpart of the famous $\phi^4$-theory
of equilibrium~\cite{JANSSEN_13}.
Rescaling the fields 
\begin{equation}
{\tilde n}({\underline x},t) \; = \; \mu \, 
{\tilde s}({\underline x},t) \, , \quad\quad
{n}({\underline x},t) \; = \; \mu^{-1} \, 
{s}({\underline x},t) \, 
\label{eq:rescal_field}
\end{equation}
the functional $S$ is invariant under the duality
transformation (so-called rapidity reversal in Reggeon field 
theory)
\begin{equation}
{\tilde s}({\underline x},t) \, \longleftrightarrow
\, -\, {s}({\underline x},-t) \, 
\label{eq:rapidity_trans}
\end{equation}
for $\mu^2=2u/ \kappa$. Note that $\mu$ is a redundant variable
from the renormalization group point of view~\cite{JANSSEN_1,JANSSEN_4}.
The rapidity reversal~(\ref{eq:rapidity_trans}) is the 
characteristic symmetry of the universality class of
directed percolation.
It is worth to  reemphasize that the rapidity reversal is obtained
from a field theoretical treatment.
Thus all models that belong to the universality class
of directed percolation obey, at least asymptotically,
the rapidity reversal after a corresponding coarse grained
procedure~\cite{JANSSEN_1}.
In other words, the rapidity symmetry may not be represented 
in the microscopic models. 
For example, 
bond directed percolation~\cite{HINRICHSEN_1} 
and site directed percolation~\cite{LUEB_35} obey the 
rapidity reversal microscopically whereas e.g.~the contact
process and the pair contact process do not.
Still, the latter two models do so asymptotically and belong
to the DP class.

Due to the continuing improvement of computer hardware,
high accurate numerical data of DP have become available 
in the last years, resulting in a fruitful and 
instructive interplay
between numerical investigations and renormalization
group analyses.
In particular, investigations of the scaling behavior of the equation of
state and of the susceptibility~\cite{JANSSEN_2,JANSSEN_3,LUEB_28}, 
of finite-size scaling functions~\cite{JANSSEN_9,LUEB_33}, 
and of dynamical scaling functions~\cite{JANSSEN_3,GRASSBERGER_P2004}
yield an impressive agreement between field theoretical and
numerical results.
For example, the universal amplitude ratio of the susceptibility 
has been calculated via $\epsilon$-expansions~\cite{JANSSEN_2}.
In second order of $\epsilon$, the error of the field theory 
estimate is about~$6\%$ for 
$D=2$~(see ref.\,\cite{LUEB_35} for a detailed discussion).
It is instructive to compare this result to the equilibrium
situation.
The corresponding $\phi^4$-theory value~\cite{NICOLL_1} differs from the
exact value of the two-dimensional Ising model~\cite{BAROUCH_1,DELFINO_2}
by roughly $115\%$.
Thus, in contrast to the $\phi^4$-theory 
the DP field theory provides excellent numerical estimates of certain
universal quantities.

In our previous works we
investigated the steady state critical scaling behavior of 
two one-dimensional models~\cite{LUEB_27}
and of one model in various dimensions~\cite{LUEB_28}, 
respectively.
In this work we extend our investigations and consider
the steady state and dynamical scaling behavior 
of five different models in various dimensions.
All lattice models are expected to belong to the universality
class of directed percolation.
So far, most works focus on the determination of the
critical exponents only, neglecting the determination
of universal scaling functions.
It turns out that checking the universality class it is 
often a more exact test to consider scaling functions 
rather than the values of the critical exponents.
While for the latter ones the variations between 
different universality classes are often small, 
the scaling functions may differ significantly~\cite{LUEB_32_BEM}.
Thus the agreement of universal scaling functions
provides not only additional but also independent and more convincing 
evidence in favor of the conjecture that the phase
transitions of two models belong to the same universality
class.
Additionly to the critical exponents and scaling functions,
universality classes are also characterized by certain
amplitude combinations (see e.g.~\cite{PRIVMAN_2}).
But these amplitude combinations are merely particular
values of the scaling functions and will be neglected in this work.

\section{Lattice models of directed percolation}
\label{sec:mod_sim}

In the following we consider various lattice 
models that belong to the universality class of directed
percolation.
First, we revisit the contact process that is well
known in the mathematical literature (see e.g.~\cite{LIGGETT_1}).
Second, we consider the Domany-Kinzel cellular 
automaton~\cite{DOMANY_1} which is very useful
in order to perform numerical investigations
of directed bond and directed site percolation.
Third, we consider the pair contact process~\cite{JENSEN_2}
that is characterized in contrast to the other models 
by infinitely many absorbing states. 
Unlike the first two models the universal scaling behavior of 
the pair contact process is still a matter of discussions 
in the literature.
Furthermore, we briefly discuss the threshold transfer 
process~\cite{MENDES_1} as well as the 
Ziff-Gulari-Barshad model~\cite{ZIFF_1}.
The latter one mimics the catalysis of 
carbon monoxide oxidation.

\subsection{Contact process}
\label{subsec:cp}

The contact process (CP) is
a continuous-time Markov
process that is usually defined on a $D$-dimensional
simple cubic lattice
(see~\cite{MARRO_1} and references therein).
A lattice site  may be empty ($n=0$) 
or occupied ($n=1$) by a particle and the 
dynamics is characterized by spontaneously occurring
processes, taking place with certain transition rates.
In numerical simulations the asynchronous 
update
is realized by a random sequential update 
scheme:~A particle 
on a randomly selected lattice site~${i}$ 
is annihilated with rate one, 
whereas particle creation takes places on an 
empty neighboring 
site with rate $\lambda N / 2 D$, i.e.,
\begin{eqnarray}
\label{eq:trans_rate_cp_annih}
n_{\scriptscriptstyle i} = 1 \quad  
&  \mathop{\longrightarrow}\limits_{1} &
\quad n_{\scriptscriptstyle i} = 0 \, ,\\
\label{eq:trans_rate_cp_creat}
n_{\scriptscriptstyle i} = 0 \quad  
&  \mathop{\longrightarrow}\limits_{\lambda N / 2 D} & 
\quad n_{\scriptscriptstyle i} = 1 \, ,
\end{eqnarray}
where~$N$ denotes the number of occupied neighbors
of~$n_{\scriptscriptstyle i}$.
Note that the rates are defined as transition probabilities
per time unit, i.e., they may be larger than one.
Thus, rescaling the time will change the transition rates.
In simulations a discrete time formulation of the
contact process is performed.
In that case a particle creation takes place at a randomly 
chosen neighbor site with probability $p=\lambda/(1+\lambda)$
whereas particle annihilation occurs with 
probability $1-p=1/(1+\lambda)$.
In dynamical simulations, the time increment $1/N_{\scriptscriptstyle \mathrm{a}}$ is
associated with each attempted elementary update 
step,
where $N_{\scriptscriptstyle \mathrm{a}}$ denotes the number of active sites.
It is usual to present the critical value in terms
of $\lambda_{\scriptscriptstyle \mathrm{c}}$ instead 
of~$p_{\scriptscriptstyle \mathrm{c}}$.

Similar to equilibrium phase transitions, it is often 
possible for absorbing phase transitions to apply an 
external field~$h$ that is conjugated
to the order parameter, i.e., to the density of active 
sites~$\rho_{\scriptscriptstyle \mathrm a}$.
Being a conjugated field it has to destroy the absorbing
phase, it has to be independent of the control parameter,
and the corresponding linear response function
has to diverge at the critical point
\begin{equation}
\chi \; = \; \frac{\partial \, 
\rho_{\scriptscriptstyle \mathrm a}}{\partial \, h} 
 \; 
\longrightarrow \; \infty \, .
\label{eq:lin_resp_apt}
\end{equation}
In case of the CP, the conjugated field is implemented
by a spontaneous creation of particles, 
i.e., the external field creates a particle at an 
empty lattice site with rate~$h$.
Clearly spontaneous particle generation destroys 
the absorbing state and therefore the absorbing phase
transition at all.
Incorporating the conjugated field, 
a series of opportunities is offered to compare
renormalization group results to those of numerical
investigations.
For example, simulations performed for non-zero field
include the measurements of the equation of state~\cite{LUEB_27},
of the susceptibility~\cite{LUEB_28}, as well as of a modified 
finite-size scaling analysis 
appropriate for absorbing phase transitions~\cite{LUEB_33,LUEB_23}.

\subsection{Domany-Kinzel automaton}
\label{subsec:dk}

An important $1+1$-dimensional stochastic cellular automaton 
exhibiting directed percolation scaling behavior is 
the Domany-Kinzel (DK) automaton~\cite{DOMANY_1}.
It is defined on a diagonal square lattice with a discrete time variable
and evolves by parallel update
according to the following rules:~A site at 
time $t$ is occupied with 
probability~$p_{\scriptscriptstyle 2}$ 
($p_{\scriptscriptstyle 1}$) if both (only one) backward
sites (at time $t-1$) are occupied.
Otherwise the site remains empty.
If both backward sites are empty a 
spontaneous particle creation takes place with 
probability $p_{\scriptscriptstyle 0}$.
Similar to the contact process, the spontaneous 
particle creation destroys the absorbing phase (empty lattice) 
and corresponds therefore to the conjugated field~$h$.

It is straight forward to generalize the $1+1$-dimensional
Domany-Kinzel automaton to higher 
dimensions (see e.g.~\cite{GRASSBERGER_8,GRASSBERGER_3,LUEB_28}).
In the following, we consider cellular automata on a 
$D+1$-dimensional body centered cubic (bcc) lattice where 
the time corresponds to the $[0,0,\ldots,0,1]$ direction.
A lattice site at time~$t$ is occupied with probability~$p$
if at least one of its $2^D$ backward neighboring 
sites (at time $t-1$) is occupied.
Otherwise the site remains empty.
This parameter choice corresponds to the probabilities
$p_{\scriptscriptstyle 1}=p_{\scriptscriptstyle 2}=\ldots
=p_{\scriptscriptstyle 2^D}=p$,
i.e., site-directed percolation (sDP) is considered.

\subsection{Pair contact process}
\label{subsec:pcp}

The pair contact process (PCP) was introduced 
by Jensen~\cite{JENSEN_2} and is one of the 
simplest models with infinitely many absorbing states 
showing a continuous phase transition.
The process is defined on a $D$-dimensional cubic lattice
and an asynchronous (random sequential) update 
scheme is applied.
A lattice site may be 
either occupied ($n=1$) or empty ($n=0$).
Pairs of adjacent occupied sites, linked by an active bond, 
annihilate each other with probability~$p$ otherwise an 
offspring is created at a neighboring site provided the 
target site is empty.
The density of active bonds $\rho_{\mathrm{a}}$ is the order 
parameter of a continuous phase transition from an active 
state  
to an inactive absorbing state without particle pairs. 
Similar to the contact process and to the Domany-Kinzel
automaton a spontaneous particle creation acts as
a conjugated field~\cite{LUEB_27}.
Since isolated particles remain inactive, any
configuration containing only isolated particles is 
absorbing.
In case of the $1+1$-dimensional pair contact process
with $L$~sites and periodic boundary conditions
the number of absorbing states is asymptotically 
given by the golden mean
$N \sim (1/2+\sqrt{5}/2)^{L}$~\cite{CARLON_1}.
In the thermodynamic limit ($L\to \infty$), the pair contact process
is characterized by infinitely many absorbing states.
Due to that non-unique absorbing phase the universality
hypothesis of Janssen and Grassberger can not be applied.
Therefore, the critical behavior of the pair contact process
was intensely investigated by 
simulations (including~\cite{JENSEN_1,JENSEN_3,DICKMAN_11,DICKMAN_4,DICKMAN_5}).
It was shown numerically that the critical scaling behavior 
of the $1+1$-dimensional pair contact process 
is characterized by the same critical 
exponents~\cite{JENSEN_2,JENSEN_3} as well
as by the same universal scaling functions as directed 
percolation~\cite{LUEB_27}.
In particular the latter result provides a convincing
identification of the universal behavior.
Thus despite the different structure of the absorbing phase,
the $1+1$-dimensional pair contact process belongs to the 
directed percolation universality class.
This numerical evidence confirms a corresponding 
renormalization group analysis
predicting DP universal behavior~\cite{MUNOZ_1}
in all dimensions.
But the scaling behavior of the PCP in higher dimension is still
a matter of controversial discussions.
A recently performed renormalization group analysis conjectures 
that the pair contact process exhibits a
dynamical percolation-like
scaling behavior~\cite{WIJLAND_1,WIJLAND_3}.
A dynamical percolation cluster at criticality equals
a fractal cluster of ordinary percolation
on the same lattice.
Thus, the dynamical percolation universality 
class~\cite{GRASSBERGER_9,CARDY_3,JANSSEN_6} differs from the 
directed percolation universality class. 
In particular the upper 
critical dimension
equals \mbox{$D_{\scriptscriptstyle \mathrm{c}}=6$}
instead of \mbox{$D_{\scriptscriptstyle \mathrm{c}}=4$} for DP.
Furthermore, the dynamical scaling behavior of the PCP is
a matter of controversial discussions.
Deviations from the directed percolation behavior due to
strong corrections to scaling~\cite{DICKMAN_11} as well as
non-universal behavior of the exponents~\cite{ODOR_7} are observed
in low dimension systems.

So far, the investigations of the PCP are limited to the
$1+1$-dimensional~\cite{LUEB_27,JENSEN_1,JENSEN_3,DICKMAN_11,DICKMAN_4,DICKMAN_5} 
and $2+1$-dimensional~\cite{SILVA_4} systems.
In this work we consider for the first time the PCP
in higher dimensions and identify the scaling behavior
via universal scaling functions.

\subsection{Threshold transfer process}
\label{subsec:ttp}

The threshold transfer process (TTP) was
introduced in~\cite{MENDES_1}.
Here, lattice sites may be empty ($n=0$), 
occupied by one particle ($n=1$),
or occupied by two particles ($n=2$).
Double occupied lattice sites are 
considered as active.
In that case both particles may move to the 
left ($\mathrm l$) or right ($\mathrm r$) 
neighbor if possible, i.e.,
\begin{eqnarray}
\label{eq:ttp_dynamic_rules}   
n_{\scriptscriptstyle \mathrm{l}} & \longrightarrow &
n_{\scriptscriptstyle \mathrm{l}} + 1 \quad\quad
{\mathrm{if}} \quad n_{\scriptscriptstyle \mathrm{l}}<2 \, ,\nonumber \\
n_{\scriptscriptstyle \mathrm{r}} & \longrightarrow &
n_{\scriptscriptstyle \mathrm{r}} + 1 \quad\quad
{\mathrm{if}} \quad n_{\scriptscriptstyle \mathrm{r}}<2 \, .
\end{eqnarray}
Additionally to the particle movement, creation and annihilation
processes are incorporated.
A particle is created at an empty lattice site ($0 \to 1$) 
with probability~$r$ whereas a particle annihilation
($1\to 0$) takes place with probability $1-r$.
In the absence of double occupied sites the dynamics 
is characterized by a fluctuating steady state with a 
density~$r$ of single occupied sites.
The density of double occupied sites is identified as the
order parameter of the process, and any configuration
devoid of double occupied sites is absorbing.
The probability~$r$ controls the particle density,
and a non-zero density of active sites occurs only
for sufficiently large values of~$r$.
In contrast to the infinitely many frozen absorbing 
configurations of the pair contact process, the
threshold transfer process is characterized by 
fluctuating absorbing states.
Nevertheless steady state numerical simulations of 
the $1+1$-dimensional threshold transfer process 
yield critical exponents that are in agreement 
with the corresponding DP values~\cite{MENDES_1}.
So far, no systematic analysis of the TTP in higher
dimensions was performed.

In this work we limit our investigations to
the $2+1$-dimensional TTP.
Analogous to the $1+1$-dimensional case, 
both particles of a given active site are tried to transfer
to randomly chosen empty or single
occupied nearest neighbor sites.
Furthermore, we apply an external field that is conjugated
to the order parameter.
In contrast to the models discussed above the conjugated field
can not be implemented by particle creation.
Particle creation with rate $h$ would affect
the particle density, i.e., the control parameter of the
phase transition.
But the conjugated field has to be independent of the 
control parameter.
Therefore, we implement the conjugated field by a diffusion-like
field that acts by particle movements.
A particle on a given lattice site moves to a randomly 
selected neighbor with probability~$h$, if $n<2$.
Thus the conjugated field of the TTP differs from the
conjugated field of the Domany-Kinzel automaton, the contact 
process, and the pair contact process.

\subsection{Ziff-Gulari-Barshad model}
\label{subsec:zgb}

Another model exhibiting a directed percolation-like
absorbing phase transition is 
the
Ziff-Gulari-Barshad
(ZGB) model~\cite{ZIFF_1}.
This model mimics the heterogeneous catalysis of the
carbon monoxide oxidation 
\begin{equation}
2\, {\mathrm {CO}} \; + \; {\mathrm O}_{\scriptscriptstyle 2} \;
\longrightarrow \; 2\, {\mathrm {CO}}_{\scriptscriptstyle 2} 
\label{eq:zgb_CO_O2_2CO2}
\end{equation}
on a catalytic material, e.g.~platinum.
The catalytic surface is represented by a square lattice
where ${\mathrm {CO}}$ or ${\mathrm {O}}_{\scriptscriptstyle 2}$ can
be adsorbed from a gas phase with concentration $y$ for 
carbon monoxide and $1-y$ for oxygen, respectively.
The concentration $y$ is the control parameter of the 
model determining the density of the components 
on the catalytic surface.
Adsorbed oxygen molecules  
dissociate at the catalytic surface into pairs
of ${\mathrm {O}}$~atoms.
It is assumed that the lattice sites are either
empty, occupied by a ${\mathrm {CO}}$ molecule,
or occupied by an ${\mathrm {O}}$ atom.
Adjacent ${\mathrm {CO}}$ and ${\mathrm {O}}$ react
instantaneously and the resulting ${\mathrm {CO}}_{\scriptscriptstyle 2}$
molecule leaves the system.
Obviously, the system is trapped in absorbing configurations
if the lattice is completely covered by carbon monoxide
or completely covered by oxygen.
The dynamics of the system is attracted by these absorbing
configurations for sufficiently large ${\mathrm {CO}}$ concentrations
and for sufficiently large ${\mathrm {O}}_{\scriptscriptstyle 2}$ concentrations.
Numerical simulations show that catalytic activity occurs 
in the range \mbox{$0.390 \lesssim y \lesssim 0.525$}~\cite{JENSEN_8} 
only.
The system undergoes a second order phase
transition to the oxygen passivated phase whereas
a first order phase transition takes place if the
${\mathrm {CO}}$ passivated phase is approached.
In particular, the continuous phase transition  
is expected to belong to the universality class 
of directed percolation~\cite{GRINSTEIN_3}.
This conjecture is supported by numerical determinations 
of certain critical exponents~\cite{JENSEN_8,VOIGT_1}.
At first glance, it might be surprising 
that the ZGB model exhibits directed percolation like behavior
since the ZGB model is characterized by two
distinct chemical components, ${\mathrm {CO}}$
and ${\mathrm {O}}$.
But the catalytic activity is connected
to the density of vacant sites, i.e., to a
single component order parameter~\cite{GRINSTEIN_3}.
Thus the observed directed percolation exponents are in 
full agreement with the universality hypothesis of Janssen 
and Grassberger.
But one has to stress that the ZGB model is an 
oversimplified representation of the catalytic 
carbon monoxide oxidation.
A more realistic modeling has to incorporate for
example mobility and desorption processes
as well as a repulsive interaction between 
adsorbed oxygen molecules (see e.g.~\cite{ZIFF_1,LIU_1}).
These features affect the critical behavior and
drive the experimental system out of the directed percolation
universality class.

In this work, we focus on the two-dimensional
ZGB model and determine certain scaling functions
for the first time.
Therefore, we apply an external field conjugated to
the order parameter.
Since the order parameter is connected to the density of vacant
sites of the catalytic reaction, the conjugated field
could be implemented via a desorption rate~$h$ of adsorbed 
oxygen molecules.\\

In summary, we investigate the scaling behavior
of five different models
spanning a broad range of interaction details, such
as different update schemes (random sequential as well as 
parallel update),
different lattice structures (simple cubic and bcc lattice types),
different inactive backgrounds 
(trivial, fluctuating or quenched background of inactive
particles),
different structures of the absorbing phase 
(unique absorbing state or infinitely many absorbing states),
as well as different implementations of the 
conjugated field (implemented via particle
creation, particle diffusion or particle desorption).
Nevertheless we will see that all models are characterized by the 
same scaling behavior, i.e., they belong to the 
same universality class.

\section{Steady state scaling behavior}
\label{sec:steady_state_scaling}

\begin{figure}[t]
\centering
  \includegraphics[width=8.6cm,angle=0]{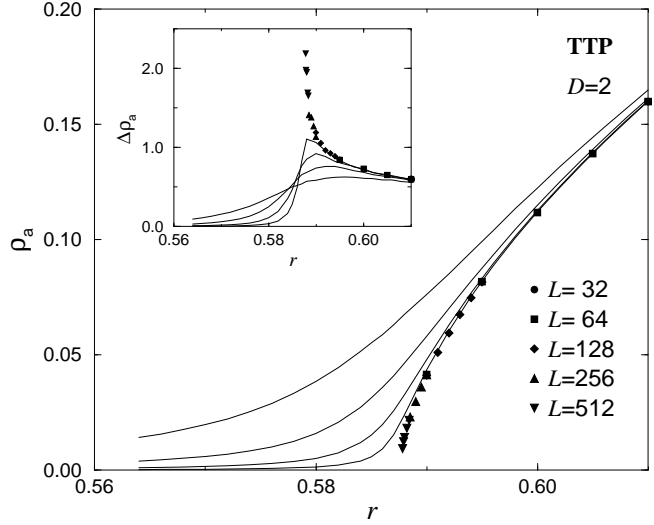}
  \caption{
The order parameter~$\rho_{\scriptscriptstyle \mathrm a}$
and its fluctuations~$\Delta\rho_{\scriptscriptstyle \mathrm a}$ 
(inset) of the $2+1$-dimensional
transfer threshold process (TTP) on a square lattice for various
values of the field (from $h=10^{-5}$ to $h=10^{-3}$).
The symbols mark the zero-field behavior.
The data are obtained from simulations on various system
sizes with periodic boundary conditions.}
  \label{fig:ttp_op_fluc_01} 
\end{figure}

In this section we consider the steady state scaling behavior 
close to the transition point.
Therefore, we performed steady state simulations of the five
models described above.
In particular, we consider the density of active sites
$\rho_{\scriptscriptstyle \mathrm a}=
\langle L^{-\scriptscriptstyle D} 
N_{\scriptscriptstyle \mathrm a} \rangle$, i.e., the
order parameter as a function of the control parameter
and of the conjugated field.
Analogous to equilibrium phase transitions, the
conjugated field results in a rounding of the zero-field
curves and the order parameter behaves smoothly as a function 
of the control parameter for finite field values 
(see figure\,\ref{fig:ttp_op_fluc_01}).
For $h\to 0$ we recover the non-analytical 
order parameter behavior.
Additionally to the order parameter, we investigate the
order parameter 
fluctuations $\Delta\rho_{\scriptscriptstyle \mathrm a}
=L^{\scriptscriptstyle D}(
\langle \rho_{\scriptscriptstyle \mathrm a}^{\scriptscriptstyle 2} \rangle -
\langle \rho_{\scriptscriptstyle \mathrm a} \rangle^{\scriptscriptstyle 2} 
)$
and its susceptibility~$\chi$.
The susceptibility is obtained by performing the numerical 
derivative of the order parameter 
$\rho_{\scriptscriptstyle \mathrm a}$ with 
respect to the conjugated field (\ref{eq:lin_resp_apt}).
Similar to the equilibrium phase transitions, the fluctuations
and the susceptibility display a characteristic peak at the 
critical point.
In the limit $h\to 0$ this peak diverges,
signalling the critical point.

In the following, we take only those simulation data
into account
where the (spatial) correlation length $\xi_{\perp}$ 
is small compared to the system size~$L$.
In that case, the order parameter,
its fluctuations, as well as the order parameter susceptibility
can be described by the following generalized homogeneous functions
\begin{eqnarray}
\label{eq:scal_ansatz_EqoS}
\rho_{\scriptscriptstyle \mathrm a}(\delta p, h) 
\; & \sim & \; 
\lambda^{-\beta}\, \, {\tilde R}
(a_{\scriptscriptstyle p}  
\delta p \; \lambda, a_{\scriptscriptstyle h} h \;
\lambda^{\sigma}) \, ,\\[2mm]
\label{eq:scal_ansatz_Fluc}
a_{\scriptscriptstyle \Delta} \,
\Delta \rho_{\scriptscriptstyle \mathrm a}(\delta p, h) 
\; & \sim & \; 
\lambda^{\gamma^{\prime}}\, \, {\tilde D}
(a_{\scriptscriptstyle p} \delta p \; \lambda, 
a_{\scriptscriptstyle h} h \, \lambda^{\sigma})  \, , \\[2mm]
\label{eq:scal_ansatz_Susc}
a_{\scriptscriptstyle \chi} \,
\chi(\delta p, h) 
\; & \sim & \; 
\lambda^{\gamma}\, \, {\tilde {\mathrm{X}}}
(a_{\scriptscriptstyle p} \delta p \; \lambda, 
a_{\scriptscriptstyle h} h \, \lambda^{\sigma})  \, ,
\end{eqnarray}
with the order parameter exponent~$\beta$, 
the field exponent~$\sigma$ (corresponding to the gap exponent
in equilibrium),
the fluctuation exponent~$\gamma^{\prime}$,
and the susceptibility exponent~$\gamma$.
Here, $h$ denotes the conjugated field
and $\delta p$ denotes the distance to the critical point,
e.g.~$\delta p = (\lambda - \lambda_{\scriptscriptstyle \mathrm c})
/\lambda_{\scriptscriptstyle \mathrm c}$ 
for the contact process,
$\delta p = (p-p_{\scriptscriptstyle \mathrm c})/p_{\scriptscriptstyle \mathrm c}$ 
for site-directed percolation and for the pair contact process, 
$\delta p = (r - r_{\scriptscriptstyle \mathrm c})/r_{\scriptscriptstyle \mathrm c}$ 
for the threshold transfer process, etc..
The so-called non-universal metric factors 
$a_{\scriptscriptstyle p}$, $a_{\scriptscriptstyle h}$, 
$a_{\scriptscriptstyle \Delta}$
and $a_{\scriptscriptstyle \chi}$
contain all non-universal system dependent features~\cite{PRIVMAN_3}
(e.g.~the lattice structure, the range of interaction,
the used update scheme,
as long as the interaction
decreases sufficient rapidly as a function of separation, etc.).
Once the non-universal metric factors 
are chosen in a specified way (see below), 
the universal scaling functions ${\tilde R}$, 
${\tilde D}$, and ${\tilde {\mathrm{X}}}$ 
are the same for all systems within a given universality class.
The above scaling forms are valid for $D\neq D_{\scriptscriptstyle \mathrm c}$.
At the upper critical dimension $D_{\scriptscriptstyle \mathrm c}$
they have to be modified by logarithmic corrections~\cite{LUEB_26,JANSSEN_3}.

\begin{figure}[b]
\centering
  \includegraphics[width=8.6cm,angle=0]{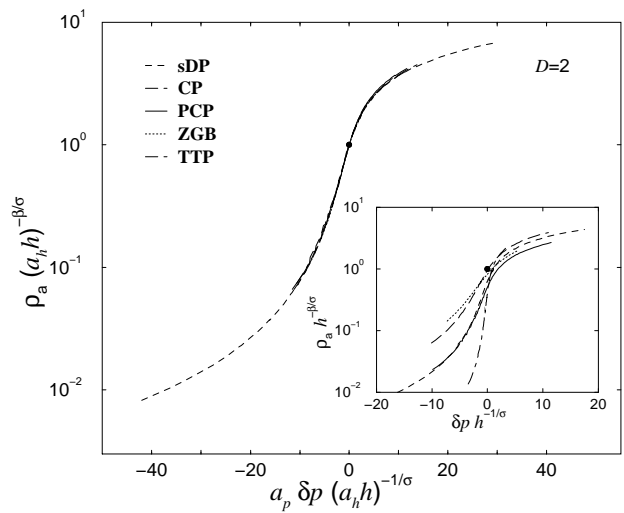}
  \caption{
The universal scaling function ${\tilde R}(x,1)$ of the
directed percolation universality class.
The data are plotted according 
to equation\,(\ref{eq:scal_ansatz_EqoS_collapse}).
All models considered are characterized by the same universal
scaling function, an impressive 
manifestation of the robustness of the directed 
percolation universality class with respect to 
variations of the microscopic interactions.
Neglecting the non-universal metric factors 
$a_{\scriptscriptstyle {p}}$ 
and $a_{\scriptscriptstyle h}$ each model
is characterized by its own scaling function (see inset).
For all models the scaling plots contain at least four
different curves corresponding to four different
field values (see e.g.~figure\,\protect\ref{fig:ttp_op_fluc_01}).
The circles mark the condition ${\tilde R}(0,1)=1$.
   }
  \label{fig:uni_dp_eqos_HS_2d} 
\end{figure}

Throughout this work we norm the universal scaling functions by 
${\tilde R}(1,0) =  1$, 
${\tilde R}(0,1)  = 1$, and 
${\tilde D}(0,1)  = 1$.
In that way, the non-universal metric factors 
$a_{\scriptscriptstyle p}$, $a_{\scriptscriptstyle h}$, and 
$a_{\scriptscriptstyle \Delta}$
are determined by the amplitudes of the power-laws
\begin{eqnarray}
\label{eq:metric_factors_a_rho}
\rho_{\scriptscriptstyle \mathrm a}(\delta p, h=0) \; & \sim & \; 
(a_{\scriptscriptstyle p} \, \delta p)^{\beta} \, ,\\
\label{eq:metric_factors_a_h}   
\rho_{\scriptscriptstyle \mathrm a}(\delta p =0, h) \; & \sim & \; 
(a_{\scriptscriptstyle h} \, h)^{\beta / \sigma} \, , \\
\label{eq:metric_factors_a_Delta}   
a_{\scriptscriptstyle \Delta} \,
\Delta\rho_{\scriptscriptstyle \mathrm a}(\delta p=0, h) \; & \sim &\; 
(a_{\scriptscriptstyle h} \, h)^{-\gamma^{\prime}/\sigma} \, .
\end{eqnarray}
Taking into consideration that the susceptibility is defined as the
derivative of the order parameter with respect to the 
conjugated field~(\ref{eq:lin_resp_apt})
we find 
\begin{equation}
{\tilde {\mathrm X}}(x,y) \; = \; \partial_y \, {\tilde R}(x,y) \, ,
\quad \quad \quad
a_{\scriptscriptstyle \chi}\; = \; a_{\scriptscriptstyle h}^{-1} \, ,
\label{eq:sus_func_metric_factor}
\end{equation}
as well as the scaling law
\begin{equation}
\gamma \; = \; \sigma  \, - \, \beta \, .
\label{eq:widom_apt}
\end{equation}
This scaling law corresponds to the well 
known Widom
law of equilibrium phase transitions.
Furthermore, comparing equation\,(\ref{eq:scal_ansatz_Susc})
for $\delta p=0$ to the definition of the susceptibility
\begin{equation}
a_{\scriptscriptstyle \chi} \,
\chi(\delta p, h) 
\; \sim  \; (a_{\scriptscriptstyle h} h)^{-\gamma/\sigma}\;
{\tilde {{\mathrm{X}}}}(0,1) \, ,
\quad
\chi \; = \; \partial_{\scriptscriptstyle h} \, \rho_{\scriptscriptstyle \mathrm a} 
\; = \; \partial_{\scriptscriptstyle h} \, (a_{\scriptscriptstyle h} h)^{\beta/\sigma}
\end{equation}   
leads to 
\begin{equation}
{\tilde {\mathrm{X}}}(0,1) \; = \; \frac{\,\beta\,}{\sigma} \, .
\label{eq:susc_Chi_01}
\end{equation}
This result offers a useful consistency check of 
the numerical estimates of the susceptibility.
Furthermore, it is worth mentioning that the validity of the scaling
form (\ref{eq:scal_ansatz_EqoS}) implies the required
singularity of the susceptibility (\ref{eq:lin_resp_apt}), i.e.,
it confirms that the applied external field is
conjugated to the order parameter.

\begin{figure}[t]
\centering
  \includegraphics[width=8.6cm,angle=0]{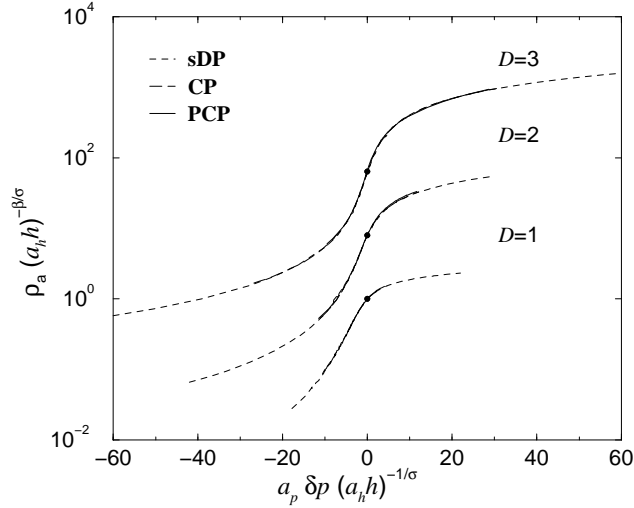}
  \caption{
The universal scaling function ${\tilde R}(x,1)$ of the
directed percolation universality class in various
dimensions.
The two- and three-dimensional data are vertically shifted
by a factor $8$ and $64$ in order to avoid overlaps.
The circles mark the condition ${\tilde R}(0,1)=1$.
   }
  \label{fig:uni_dp_eqos_HS_123d} 
\end{figure}

Choosing $a_{\scriptscriptstyle h} h \, \lambda^{\sigma}=1$ in
equations~(\ref{eq:scal_ansatz_EqoS}-\ref{eq:scal_ansatz_Susc}) 
we obtain the scaling forms
\begin{eqnarray}
\label{eq:scal_ansatz_EqoS_collapse}
\rho_{\scriptscriptstyle \mathrm a}(\delta p, h) 
\; & \sim & \; 
(a_{\scriptscriptstyle h}\, h)^{\beta/\sigma}\, \, {\tilde R}
(a_{\scriptscriptstyle p}  
\delta p \; (a_{\scriptscriptstyle h}\, h)^{-1/\sigma}, 1) \, ,\\[2mm]
\label{eq:scal_ansatz_Fluc_collapse}
a_{\scriptscriptstyle \Delta} \,
\Delta \rho_{\scriptscriptstyle \mathrm a}(\delta p, h) 
\; & \sim & \; 
(a_{\scriptscriptstyle h}\, h)^{-\gamma^{\prime}/\sigma}\, \, {\tilde D}
(a_{\scriptscriptstyle p}  
\delta p \; (a_{\scriptscriptstyle h}\, h)^{-1/\sigma}, 1) \, ,\\[2mm]
\label{eq:scal_ansatz_Susc_collapse}
a_{\scriptscriptstyle \chi} \,
\chi(\delta p, h) 
\; & \sim & \; 
(a_{\scriptscriptstyle h}\, h)^{-\gamma/\sigma}\, \, {\tilde {\mathrm{X}}}
(a_{\scriptscriptstyle p}  
\delta p \; (a_{\scriptscriptstyle h}\, h)^{-1/\sigma}, 1) \, .
\end{eqnarray}
Thus plotting the rescaled quantities
\begin{equation}
\rho_{\scriptscriptstyle \mathrm a} \; 
(a_{\scriptscriptstyle h} \, h)^{-\beta/\sigma} \, , \quad
a_{\scriptscriptstyle \Delta} \, \Delta \rho_{\scriptscriptstyle \mathrm c}
\; (a_{\scriptscriptstyle h}\, h)^{\gamma^{\prime}/\sigma} \, ,
\quad
a_{\scriptscriptstyle \chi} \, \chi \;
(a_{\scriptscriptstyle h}\, h)^{-\gamma/\sigma} \, 
\label{eq:rescal_quantities_steady_state}
\end{equation}
as a function of the rescaled control parameter
$x=a_{\scriptscriptstyle p} \delta p \, (a_{\scriptscriptstyle h} h)^{-1/\sigma}$
the data of all systems belonging to the
same universality class have to collapse onto the universal 
curves ${\tilde R}(x,1)$, ${\tilde D}(x,1)$, 
and ${\tilde {\mathrm{X}}}(x,1)$.
This is shown for ${\tilde R}(x,1)$
in figure~\ref{fig:uni_dp_eqos_HS_2d}
where the rescaled order parameter is plotted
for various two-dimensional models.
For this purpose the best known estimates for the critical
exponents, as given in Table\,\ref{table:dp_exponents}, are used.
As can be seen the data of all considered models
collapse onto the same scaling function, 
clearly supporting the assumption that all models
belong to the universality class of directed percolation.
Furthermore, the data-collapse confirms the accuracy of the
numerically estimated values~\cite{VOIGT_1} of the critical exponents.

To limit the numerical effort we skip in the following
analysis the TTP and ZGB and focus to the scaling behavior
of the CP, sDP, and the PCP in various dimension.
The universal scaling function ${\tilde R}(x,1)$ is
displayed in figure~\ref{fig:uni_dp_eqos_HS_123d} for $D=1,2,3$.
For each dimension, the data of the three models
collapse onto the unique scaling function.
As expected the scaling functions vary with 
the spatial dimensionality below
for $D<D_{\scriptscriptstyle \mathrm c}$.

\begin{figure}[t]
\centering
  \includegraphics[width=8.6cm,angle=0]{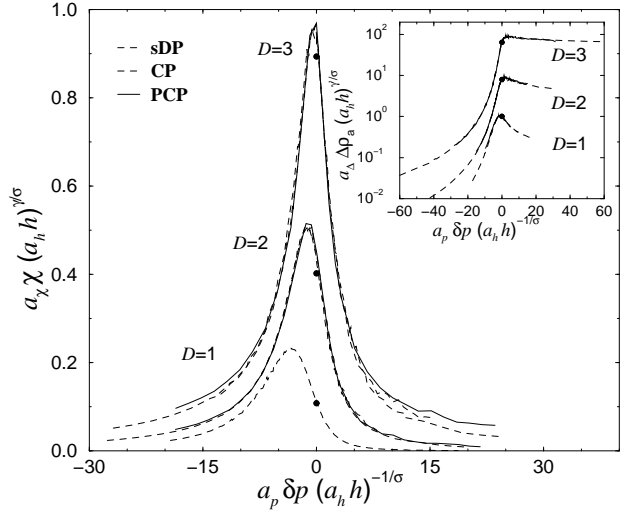}
  \caption{
The universal scaling functions 
of the susceptibility ${\tilde {\mathrm{X}}}(x,1)$
and of the fluctuations
${\tilde D}(x,1)$ (inset) in various dimensions.
In case of the susceptibility the 
two- and three-dimensional data
are vertically shifted
by a factor $3/2$ and $9/4$ in order to avoid overlaps.
The circles mark the condition ${\tilde {\mathrm{X}}}(0,1)=\beta/\sigma$
and reflect the accuracy of the performed analysis.
The two- and three-dimensional fluctuation 
data are vertically shifted
by a factor $8$ and $64$ in order to avoid overlaps.
The circles mark the condition ${\tilde D}(0,1)=1$.
   }
  \label{fig:uni_dp_fluc_susc_123d} 
\end{figure}

The universal scaling functions of the order parameter
fluctuations and the order parameter susceptibility 
are shown in figure~\ref{fig:uni_dp_fluc_susc_123d}.
The susceptibility is obtained by performing 
the numerical derivative
of the order parameter with respect to the 
conjugated field.
The perfect data-collapses confirm the scaling
forms (\ref{eq:scal_ansatz_Fluc},\ref{eq:scal_ansatz_Susc}).
All scaling functions exhibit for $D=1,2,3$ a clear maximum 
signalling the divergence of 
$\Delta\rho_{\scriptscriptstyle \mathrm a}$ and
$\chi$ at the critical point.
The susceptibility data fulfill equation~(\ref{eq:susc_Chi_01}),
reflecting the accuracy of the performed analysis.
In summary, all considered models are characterized
by the same critical exponents and the same steady state 
scaling functions ${\tilde R}$, ${\tilde D}$, and
${\tilde {\mathrm{X}}}$.
Thus the steady state scaling behavior of the contact process, 
the Domany-Kinzel automaton, 
the pair contact process, 
the threshold transfer process,
as well as of the Ziff-Gulari-Barshad model
belong to the same universality class for $D=1,2,3$.

\subsection{Mean field regime}
\label{subsec:stetady_state_mf}

Mean field theories of all models considered can be simply
derived and are well established 
(see for example~\cite{KINZEL_2,TOME_1,RIEGER_1,MARRO_1,ATMAN_1,LUEB_27}).
These mean field theories present not only some insight
into the critical behavior, they become valid above the
upper critical dimension~$D_{\scriptscriptstyle \mathrm c}$.
Thus analytical expressions for the scaling
functions become available for $D>D_{\scriptscriptstyle \mathrm c}$.
In the case of directed percolation the mean-field
scaling functions are given 
by~(see e.g.~\cite{MORI_1,LUEB_28})
\begin{eqnarray}
\label{eq:uni_scal_mf_R}
{\tilde R}_{\scriptscriptstyle \mathrm {MF}}  
(x, y) & = &
\frac{x}{2} \, + \, \sqrt{y \, + \,
\left (\frac{x}{2} \right )^2 \;} \, ,   \\
\label{eq:uni_scal_mf_D}
{\tilde D}_{\scriptscriptstyle \mathrm {MF}}  
(x , y) & = &
\frac{{\tilde R}_{\scriptscriptstyle \mathrm {MF}}(x,y)}
{\,{\sqrt{y \, + \,\left ( {x}/{2} \right )^2 \;}}\,} \, , \\
\label{eq:uni_scal_mf_X}
{\tilde {\mathrm{X}}}_{\scriptscriptstyle \mathrm {MF}}  
(x , y) & = & 
\frac{1}{\,2\, \,{\sqrt{y \, + \,\left ( {x}/{2} \right )^2 \;}}\,} \, , 
\end{eqnarray}
i.e., the mean-field exponents are 
$\beta_{\scriptscriptstyle \mathrm {MF}} =1$, 
$\sigma_{\scriptscriptstyle \mathrm {MF}}=2$, 
$\gamma_{\scriptscriptstyle \mathrm {MF}}=1 $, and
$\gamma^{\prime}_{\scriptscriptstyle \mathrm {MF}} =0$.

\begin{figure}[t] 
\centering        
  \includegraphics[width=8.6cm,angle=0]{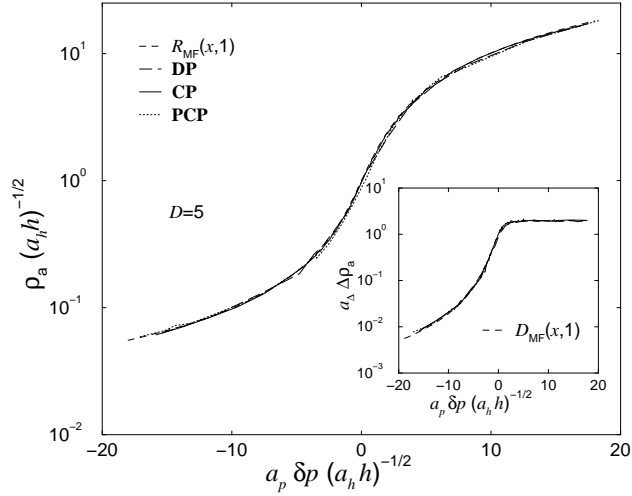}
  \caption{
The universal scaling function of the order parameter
${\tilde R}(x,1)$ and the fluctuations
${\tilde D}(x,1)$ (inset) for $D=5$.
The five-dimensional data agree with the corresponding
mean field functions (\ref{eq:uni_scal_mf_EqoS_Rx1})
and (\ref{eq:uni_scal_mf_Fluc_Dx1}), respectively.
   }
  \label{fig:uni_dp_eqos_fluc_5d} 
\end{figure}

The scaling behavior of the fluctuations deserves
comment.
The exponent $\gamma^{\prime}_{\scriptscriptstyle \mathrm {MF}} =0$
corresponds to a jump of the 
fluctuations and the scaling 
form (\ref{eq:scal_ansatz_Fluc}) reduces to
\begin{equation}
a_{\scriptscriptstyle \Delta} \; 
\Delta \rho_{\scriptscriptstyle \mathrm a}(\delta p, h) 
\; \sim \; {\tilde D}_{\scriptscriptstyle \mathrm {MF}} 
(a_{\scriptscriptstyle p} \delta p \; \lambda, 
a_{\scriptscriptstyle h} h \, \lambda^{\sigma}) \, .
\label{eq:uni_fluc_scal_fluc_mf}
\end{equation}
Using again ${\tilde D}_{\scriptscriptstyle \mathrm {MF}} (0,1)=1$,
the non-universal metric factor 
$a_{\scriptscriptstyle \Delta}$ is determined
by 
\begin{equation}
a_{\scriptscriptstyle \Delta} \; = \;
\frac{1}{\, \Delta \rho_{\scriptscriptstyle \mathrm a} 
(\delta p=0, h)\,} \, .
\label{eq:mf_fluc_d}
\end{equation}

Numerical data of the five-dimensional models are
presented in figure~\ref{fig:uni_dp_eqos_fluc_5d}.
A good data-collapse of the rescaled data 
[rescaled according to equations 
(\ref{eq:scal_ansatz_EqoS_collapse})-(\ref{eq:scal_ansatz_Susc_collapse})]
with the universal mean field scaling functions
\begin{eqnarray}
\label{eq:uni_scal_mf_EqoS_Rx1}
{\tilde R}_{\scriptscriptstyle {\mathrm {MF}}} 
(x, 1) & = &
\frac{x}{2} 
\, + \, \sqrt{1 \, + \, \left ( \frac{x}{2} \right )^2 \;} \, , \\[2mm]
\label{eq:uni_scal_mf_Fluc_Dx1}
{\tilde D}_{\scriptscriptstyle {\mathrm {MF}}} 
(x, 1) & = &  1 \, + \, 
\frac{x}
{\,2 \; \sqrt{1 \, + \, \left ( {x}/{2} \right )^2 \;}\, } \, 
\end{eqnarray}
is obtained.

The observed agreement of the data of the 
five-dimensional models with the mean field
scaling functions ${\tilde R}_{\scriptscriptstyle \mathrm {MF}}(x,1)$
and ${\tilde D}_{\scriptscriptstyle \mathrm {MF}}(x,1)$
leads to the result that the 
upper critical dimension is less than five,
in agreement with the renormalization group 
conjecture~$D_{\scriptscriptstyle \mathrm c}=4$~\cite{OBUKHOV_2,CARDY_1}.
This is a non-trivial result.
The value of $D_{\scriptscriptstyle \mathrm c}$ is often predicted by 
field theoretical treatments, and a reliable and independent
determination of the upper critical dimension 
is therefore of particular interest.
For example, two contrary field theories conjecture
for the pair contact process $D_{\scriptscriptstyle \mathrm c}=4$~\cite{MUNOZ_1}
and $D_{\scriptscriptstyle \mathrm c}=6$~\cite{WIJLAND_1,WIJLAND_3}, respectively.
The latter result is in clear contradiction to the
observed mean field behavior in $D=5$.

\subsection{Upper critical dimension}
\label{subsec:stetady_state_dc}

The scaling behavior at the upper critical dimension
is characterized by mean field power-laws
modified by logarithmic corrections.
Recent numerical investigations~\cite{AKTEKIN_1,LUEB_26,LUEB_28,GRUENEBERG_1} 
as well as analytical results~\cite{JANSSEN_3}
reveal that the concept of universal scaling functions
can also be applied to the upper critical dimension.
For example, the order parameter is expected to obey
the scaling form (all terms in leading order)
\begin{equation}
a_{\scriptscriptstyle \mathrm a}  \, \rho_{\scriptscriptstyle \mathrm a}(\delta p, h) 
\; \sim \; 
\lambda^{- \beta_{\scriptscriptstyle \mathrm {MF}}}\,  | \ln{\lambda}|^{\Lambda} 
\; \; {\tilde R}
(a_{\scriptscriptstyle p}  
\delta p \; \lambda \,  | \ln{\lambda}|^{\Pi} , 
a_{\scriptscriptstyle h} h \;
\lambda^{\sigma_{\scriptscriptstyle \mathrm {MF}}}\, | \ln{\lambda}|^{\mathrm H}) \, .
\label{eq:uni_scal_EqoS_R_dc}
\end{equation}
Greek capitals will be used in the following to denote the logarithmic
correction exponents.
According to this scaling form, the order parameter at zero field 
($h=0$) and at the critical density ($\delta p=0$) is given by
\begin{eqnarray}
\label{eq:uni_scal_OPzf_dc}
a_{\scriptscriptstyle \mathrm a}  \, \rho_{\scriptscriptstyle \mathrm a}(\delta p, h=0) 
& \sim & 
a_{\scriptscriptstyle p}  \,
\delta p \; \; | \ln{a_{\scriptscriptstyle p}  \delta p}|^{\mathrm B} 
\; \; {\tilde R}(1,0) \, , \\[2mm]
\label{eq:uni_scal_OPcp_dc}
a_{\scriptscriptstyle \mathrm a}  \, \rho_{\scriptscriptstyle \mathrm a} (\delta p=0, h) 
& \sim & 
\sqrt{a_{\scriptscriptstyle h}  h} \;
\, \left | \ln{\sqrt{a_{\scriptscriptstyle h} h}} \right |^{\Sigma} 
\; \; {\tilde R}(0,1) \, ,
\end{eqnarray}
with ${\mathrm B}=\Pi+\Lambda$ and $\Sigma={\mathrm H}/2+\Lambda$.
Similar to $D \neq D_{\scriptscriptstyle \mathrm c}$ 
the normalization ${\tilde R}(0,1)={\tilde R}(1,0)=1$ is used. 
Furthermore, the scaling behavior of the equation of state 
is given in leading order by
\begin{equation}
a_{\scriptscriptstyle \mathrm a}  \, \rho_{\scriptscriptstyle \mathrm a}(\delta p, h) 
\; \sim \; 
\sqrt{a_{\scriptscriptstyle h}  h}
\; \left | \ln{\sqrt{a_{\scriptscriptstyle h} h}} \right |^{\Sigma} 
\; \; {\tilde R}(x,1) \, ,
\label{eq:uni_scal_EqoS_dc}
\end{equation}
where $x$ denotes the scaling argument 
\begin{equation}
x \; = \; 
a_{\scriptscriptstyle p} \delta p \,
\sqrt{a_{\scriptscriptstyle h} h\,}^{\, -1} \,
\left | \ln{\sqrt{a_{\scriptscriptstyle h}  h}\,} \right |^{\Psi} 
\label{eq:uni_scal_arg_x_dc}
\end{equation}
with $\Psi=\Pi-{\mathrm H}/2={\mathrm B}-\Sigma$.

In case of directed percolation it is possible to
confirm the scaling form (\ref{eq:uni_scal_EqoS_R_dc}) 
by a renormalization group analysis~\cite{JANSSEN_3}, 
yielding in addition
the values $\Lambda=7/12$, $\Pi=-1/4$, and ${\mathrm H}=-1/2$.
Thus the scaling behavior of the equation of state
is determined by
${\mathrm B} = \Sigma  =  1/3$  and $\Psi=0$.

Similarly to the order parameter the following 
scaling form is used for the 
fluctuations~\cite{LUEB_26,LUEB_28}
\begin{equation}
 a_{\scriptscriptstyle \Delta}  \; \Delta\rho_{\scriptscriptstyle \mathrm a}(\delta p, h) 
\; \sim \; 
\lambda^{\gamma^{\prime}_{\scriptscriptstyle \mathrm {MF}}}  
\; | \ln{\lambda}|^{\mathrm K} 
\; \; {\tilde D}
(a_{\scriptscriptstyle p}  
\delta p \; \lambda \; | \ln{\lambda}|^{\Pi} , 
a_{\scriptscriptstyle h} h \;
\lambda^{-\sigma}\, | \ln{\lambda}|^{\mathrm H}) \, .
\label{eq:uni_scal_fluc_dc}
\end{equation}
Taking into account that numerical simulations show
that fluctuations remain finite at the critical point
(i.e.~$\gamma^{\prime}_{\scriptscriptstyle \mathrm {MF}}=0$
and ${\mathrm K}=0$~\cite{LUEB_28}) the scaling form
\begin{equation}
a_{\scriptscriptstyle \Delta}  \; \Delta\rho_{\scriptscriptstyle \mathrm a}(\delta p, h) 
\; \sim \; 
\; {\tilde D}(x,1)
\label{eq:uni_scal_Fluc}
\end{equation}
is obtained, where the scaling argument~$x$ is given 
by equation\,(\ref{eq:uni_scal_arg_x_dc}) with $\Psi = 0$.
The non-universal metric factor $a_{\scriptscriptstyle \Delta}$
is determined again by the condition ${\tilde D}(0,1)=1$.

\begin{figure}[t]
\centering
  \includegraphics[width=8.6cm,angle=0]{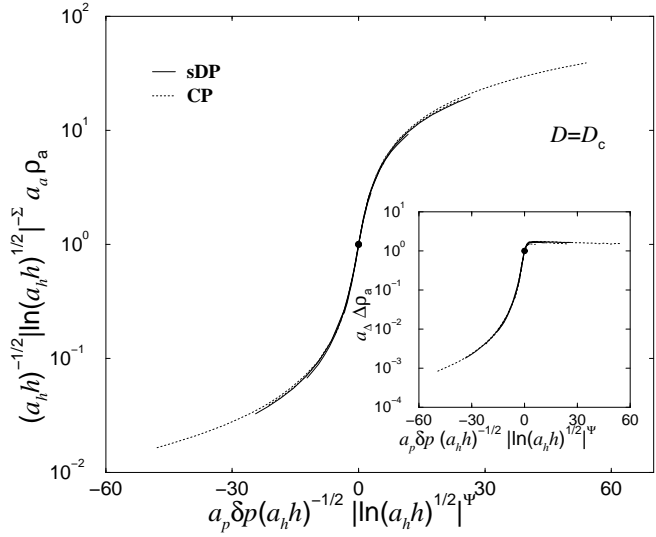}
  \caption{
The universal scaling functions
of the order parameter 
and its fluctuations (inset) 
at the upper critical dimension 
$D_{\scriptscriptstyle \mathrm c}=4$.
The logarithmic correction exponents are given by
${\mathrm B} = \Sigma = 1/3$~\protect\cite{JANSSEN_3} and $\Psi=0$.
For both models considered, the scaling plots contain at least three
different curves corresponding to three different
field values.
The circles mark the condition ${\tilde R}(0,1)=1$ 
and  ${\tilde D}(0,1)=1$, respectively.
  }
  \label{fig:uni_dp_eqos_fluc_4d} 
\end{figure}

In that way, the scaling behavior of the order parameter and 
its fluctuations at $D_{\scriptscriptstyle \mathrm c}$
is characterized by two exponents (${\mathrm B}=1/3$
and $\Sigma=1/3$) and four unknown non-universal metric factors
($a_{\scriptscriptstyle \mathrm a},a_{\scriptscriptstyle p},
a_{\scriptscriptstyle h},a_{\scriptscriptstyle \Delta}$).
The detailed analysis for sDP data is described in~\cite{LUEB_28}.
Here we present a corresponding universal scaling plot containing
data of two models, namely sDP and the CP.
The corresponding scaling plots are displayed in
Fig.\,\ref{fig:uni_dp_eqos_fluc_4d} and show
that the concept of universal scaling functions
can be applied to the upper critical dimension.

Notice that no data-collapse is obtained if logarithmic
corrections are neglected, i.e., for $\mathrm{B}=\Sigma=0$.
Thus, at least the leading logarithmic corrections have to be 
taken into account in order to study steady state
scaling functions.
It is therefore remarkable that
recently performed off-lattice simulations
of the dynamical scaling behavior at $D_{\scriptscriptstyle \mathrm c}=4$ reveal that
logarithmic corrections of higher orders
(e.g.~${\mathcal O}({\ln{\ln{t}}})$) 
are necessary to describe the numerical data~\cite{GRASSBERGER_12}.
Although the steady state results presented here are quite
convincing, we expect that even better results are obtained
by incorporation higher order corrections.

\section{Dynamical scaling behavior}
\label{sec:dynamical_scaling}

In this section we discuss the dynamical scaling
behavior close to the transition point.
We limit our attention to the so-called
activity spreading, generated from a single active seed.
Starting with the seminal work of~\cite{GRASSBERGER_4},
measurements of activity spreading have been widely 
applied in the last two decades.
In particular, they provide very accurate estimates of the
critical values~$p_{\scriptscriptstyle \mathrm c}$, 
$\lambda_{\scriptscriptstyle \mathrm c}$,\ldots 
as well as of the exponents $\delta$, $\theta$, 
and~$z$ (see below).
Here, we will focus on the scaling functions
of the survival probability~$P_{\scriptscriptstyle \mathrm a}$ and 
of the average number of active sites~$N_{\scriptscriptstyle \mathrm a}$.
At criticality both quantities obey the 
power-laws 
\begin{equation}
a_{\scriptscriptstyle P} \,    
P_{\scriptscriptstyle \mathrm a}
\; \sim  \;
(a_{\scriptscriptstyle t}  t)^{-\delta} \, , 
\quad\quad\quad\quad
a_{\scriptscriptstyle N} \, 
N_{\scriptscriptstyle \mathrm a}
\;  \sim  \; 
(a_{\scriptscriptstyle t}  t)^{\theta} \, ,
\label{eq:Pa_Na_act_spread}
\end{equation}
where $\theta$ is termed the critical initial 
slip exponent~\cite{JANSSEN_10}.
Sufficiently close to the critical point, $P_{\scriptscriptstyle \mathrm a}$ and
$N_{\scriptscriptstyle \mathrm a}$ are expected to obey the scaling forms
\begin{eqnarray}
\label{eq:scal_ansatz_Pa_dyn}   
a_{\scriptscriptstyle P} \,
P_{\scriptscriptstyle \mathrm a}(\delta p,L,t) 
& \sim  & 
\lambda^{-\delta\nu_{\parallel}}\, \, 
{\tilde P}_{\scriptscriptstyle \mathrm{pbc}}
(a_{\scriptscriptstyle p}  
\delta p \; \lambda, 
a_{\scriptscriptstyle L} L \;\lambda^{-\nu_{\perp}},
a_{\scriptscriptstyle t} t \;\lambda^{-\nu_{\parallel}}     ) \, , \\[2mm]
\label{eq:scal_ansatz_Na_dyn}   
a_{\scriptscriptstyle N} \,
N_{\scriptscriptstyle \mathrm a}(\delta p,L,t) 
& \sim  & 
\lambda^{\theta\nu_{\parallel}}\, \, {\tilde N}_{\scriptscriptstyle \mathrm{pbc}}
(a_{\scriptscriptstyle p}  
\delta p \; \lambda, 
a_{\scriptscriptstyle L} L \;\lambda^{-\nu_{\perp}},
a_{\scriptscriptstyle t} t \;\lambda^{-\nu_{\parallel}}     ) \, ,
\end{eqnarray}
where the time~$t$ and the system size~$L$ are incorporated as
additional scaling fields.
The scaling power of~$t$ has to equal the scaling power of the
correlation time $\xi_{\parallel}\propto \delta p^{-\nu_{\parallel}}$, 
whereas the scaling power
of the system size is given by the spatial correlation length 
exponent~$\nu_{\perp}$ ($\xi_{\perp}\propto \delta p^{-\nu_{\perp}}$).
In contrast to the previous section, the system size~$L$
has to be taken into account because the 
power-law behaviors (\ref{eq:Pa_Na_act_spread}) are limited by the 
finite system size.
The index $\mathrm{pbc}$ indicates that the universal
finite-size scaling functions depend on the particular
choice of the boundary conditions as well as
on the system 
shape~(see e.g.~\cite{HU_1,KANEDA_1,KANEDA_2,HUCHT_1,ANTAL_1}).
But different lattice structures are contained in the 
non-universal metric factors.

Choosing $a_{\scriptscriptstyle t} t\lambda^{-\nu_{\parallel}}=1$
the power-laws (\ref{eq:Pa_Na_act_spread}) are 
recovered
for ${\tilde P}_{\scriptscriptstyle \mathrm{pbc}}(0,\infty,1)=1$ 
as well as ${\tilde N}_{\scriptscriptstyle \mathrm{pbc}}(0,\infty,1)=1$.
The finite-size scaling forms are obtained by 
setting $a_{\scriptscriptstyle L} L \;\lambda^{-\nu_{\perp}}=1$,
yielding
\begin{eqnarray}
\label{eq:scal_ansatz_Pa_dyn_fss}      
a_{\scriptscriptstyle P} \,
P_{\scriptscriptstyle \mathrm a}(0,L,t) 
& \sim  & 
(a_{\scriptscriptstyle L} L)^{-\delta z}\, \, 
{\tilde P}_{\scriptscriptstyle \mathrm{pbc}}
(0,1,
a_{\scriptscriptstyle t} t \;
(a_{\scriptscriptstyle L} L)^{-z}     ) \, , \\[2mm]
\label{eq:scal_ansatz_Na_dyn_fss}   
a_{\scriptscriptstyle N} \,
N_{\scriptscriptstyle \mathrm a}(0,L,t) 
& \sim  & 
(a_{\scriptscriptstyle L} L)^{\theta z}\, \, 
{\tilde N}_{\scriptscriptstyle \mathrm{pbc}}
(0,1,
a_{\scriptscriptstyle t} t \;
(a_{\scriptscriptstyle L} L)^{-z}     ) \, .
\end{eqnarray}
The scaling functions 
${\tilde P_{\scriptscriptstyle \mathrm{pbc}}}(0,1,x)$ 
and ${\tilde N_{\scriptscriptstyle \mathrm{pbc}}}(0,1,x)$ 
are expected to 
decay exponentially for $t \gg t_{\scriptscriptstyle \mathrm{FSS}}$
whereas they exhibit an algebraic behavior
for $t \ll t_{\scriptscriptstyle \mathrm {FSS}}$,
with 
$t_{\scriptscriptstyle \mathrm {FSS}}=
a_{\scriptscriptstyle t}^{-1}(a_{\scriptscriptstyle L} L)^{\nu_{\perp}}$.

\begin{figure}[t] 
\centering       
  \includegraphics[width=12.0cm,angle=0]{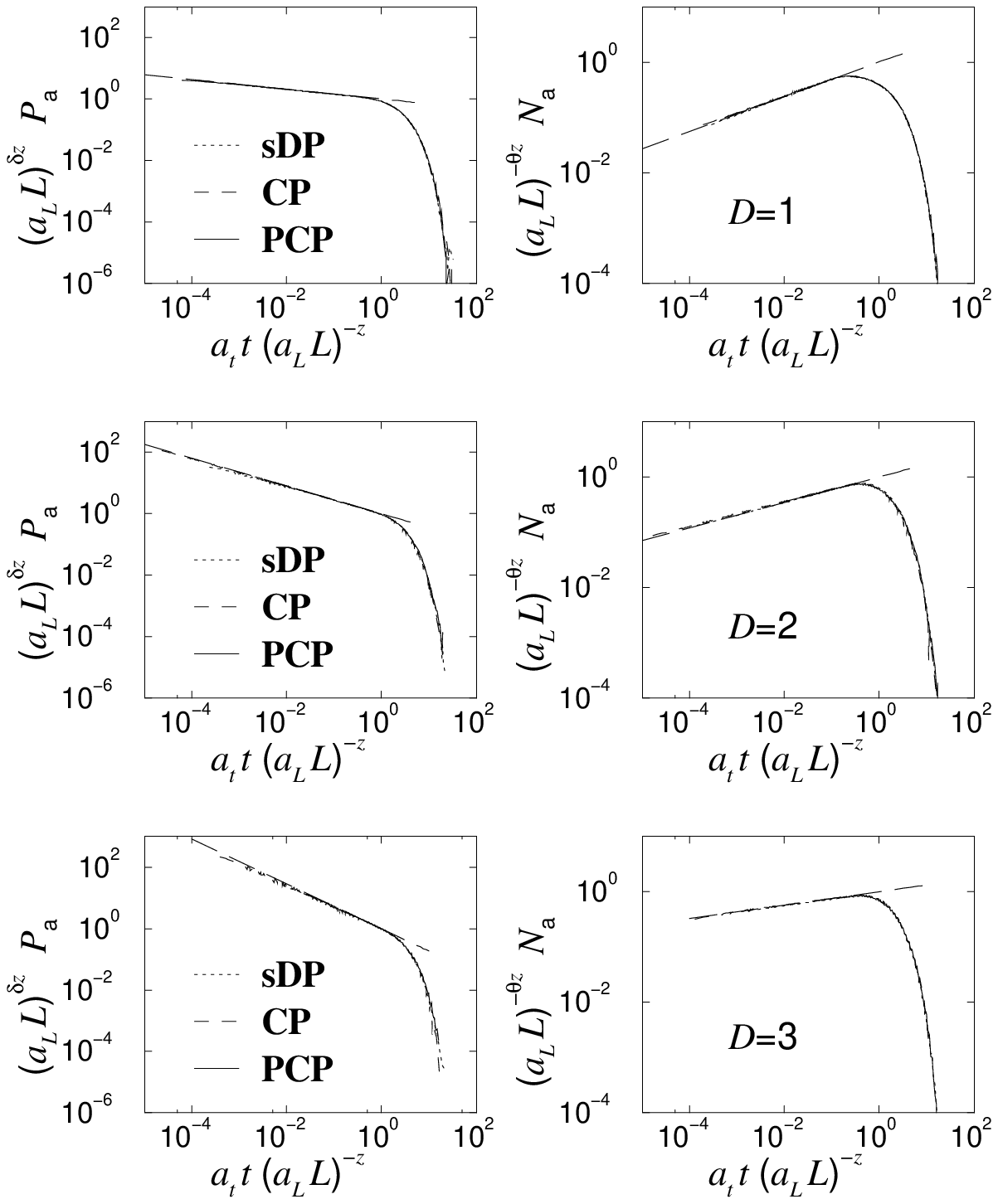}
  \caption{
The universal scaling functions 
${\tilde P}_{\scriptscriptstyle \mathrm{pbc}}$ and 
${\tilde N}_{\scriptscriptstyle \mathrm{pbc}}$
of activity spreading for various dimensions.
In case of the pair contact process the simulations
are started from a natural configuration of inactive
particles.
System sizes $L=64,128,256,512$ are considered for $D=1$,
$L=64,128,256,512$ for $D=2$, and 
$L=16,32,64,128$ for $D=3$. 
The dashed lines corresponds to the power-law
behavior of the infinite system $x^{-\delta}$ and $x^{\theta}$,
respectively.
   }
  \label{fig:uni_dp_act_spread_d123} 
\end{figure}

Performing activity spreading simulations of 
site directed percolation and of the contact process
the initial seed is implemented by a single particle on 
an empty lattice.
For absorbing phase transitions with
non-trivial absorbing states, like the pair contact process,
the scaling behavior depends upon the nature
of the initial configuration~\cite{JENSEN_3}.
In that case, spreading activity simulations have to be 
performed at the 
so-called natural density of inactive
particles~\cite{JENSEN_3,MENDES_1,LUEB_23}.
Starting with a random configuration, an absorbing state at 
criticality is prepared by the dynamics. 
An active seed is created for example by adding or moving a particle.
The resulting activity relaxation is monitored until
it ceases.
Then we start a new measurement from a random configuration and so on.
Therefore, the numerical effort is significantly increased
for systems exhibiting non-trivial absorbing states
and only small system sizes are available by simulations.
But nevertheless convincing data-collapses, including the pair contact 
process data, are obtained
and the corresponding universal scaling functions
are shown in figure\,\ref{fig:uni_dp_act_spread_d123}.
The values of the exponents used are listed 
in table\,\ref{table:dp_exponents}.
In summary, activity spreading from a localized seed
is characterized by the same universal scaling functions 
${\tilde P}_{\scriptscriptstyle \mathrm{pbc}}$ and 
${\tilde N}_{\scriptscriptstyle \mathrm{pbc}}$
for all considered models. 
Thus the CP, sDP, and especially the PCP display the same dynamical
scaling behavior at criticality.

\begin{table}[t]
\centering
\caption{
The critical exponents of directed percolation for various 
dimensions~$D$.
In $D=1$, the exponents $\gamma$, $\nu_{\protect\perp}$, and
$\nu_{\protect\parallel}$ are obtained from a 
series expansion by Jensen~\protect\cite{JENSEN_5}.
For $D=2$ and $D=3$ activity spreading simulations
are performed yielding $\delta$, $\theta$, as well 
as $z$~\protect\cite{VOIGT_1,JENSEN_6}.
Additionally, the exponent $\nu_{\protect\parallel}$ is
determined~\cite{GRASSBERGER_3,JENSEN_6} in order to 
estimate the full set of exponents via scaling laws.
}
\vspace{0.3cm}
\label{table:dp_exponents}
\begin{tabular}{|c|l|l|l|c|}
\hline
$$       
& $D=1$ {\protect{\cite{JENSEN_5}}}
& $D=2$ {\protect\cite{VOIGT_1,GRASSBERGER_3}}
& $D=3$ {\protect\cite{JENSEN_6}}
& Mean field\\  
\hline
$\beta$    &  $0.276486(8)$	& $0.5834\pm0.0030\quad$ & $0.813\pm0.009$ & $1$\\   
$\nu_{\perp}$     &  $1.096854(4)$	
& $0.7333\pm0.0075$ &  $0.584\pm0.005$& $1/2$\\    
$\nu_{\protect\parallel}$  &  $1.733847(6)$	
& $1.2950\pm0.0060$ &  $1.110\pm0.010$ & $1$\\    
$\sigma$     &  $2.554216(13)$	& $2.1782\pm0.0171$ &
$2.049\pm0.026$ & $2$\\    
$\protect\gamma^{\prime}$     &  $0.543882(16)$	&
$0.2998\pm0.0162$ & $0.126\pm0.023$& $0$\\   
$\gamma$        & $2.277730(5)$	&
$1.5948\pm0.0184$ & $1.237\pm 0.023$ & $1$\\  
$\delta$ &  $0.159464(6)$	    & $0.4505\pm0.0010$   
& $0.732\pm 0.004$ & $1$\\ 
$\theta$      &  $0.313686(8)$    & $0.2295\pm0.0010$  & $0.114\pm 0.004$ & $0$\\
$z$ 	      &  $1.580745(10)$   & $1.7660\pm0.0016$  & $1.901\pm0.005$ & $2$\\ 
\hline
\end{tabular}
\end{table}

\section{Conclusions}
\label{sec:conc}

Similar to equilibrium critical phenomena, 
the great variety of 
non-equilibrium phase transitions
can be grouped into different universality classes.
Each non-equilibrium universality class is characterized by a 
certain symmetry which is often masked 
within the Langevin equation approach, but it is
expressed clearly within the corresponding 
path integral formulation
(see e.g.~\cite{ODOR_1,JANSSEN_4} and references therein).
For example, non-equilibrium critical systems
belong to the directed percolation
universality class if the associated 
absorbing phase transition is 
described by a single component order parameter
and if the corresponding coarse grained system 
obeys the rapidity reversal symmetry (at least asymptotically).
This is clearly demonstrated in figure\,\ref{fig:uni_dp_eqos_HS_2d}
where the data of five different models is shown.
Although the models exhibit different interaction details
their rescaled data collapse onto a unique universal scaling curve.
This scaling plot is an impressive manifestation of the
robustness of the DP universality class.
On the other hand, it allows the identification of the irrelevant
parameters, i.e., those parameters which do not affect the 
universality class.
As expected, the used update scheme, the lattice structure
and the different implementation schemes of the conjugated
field do not affect the scaling behavior.
But it is worth mentioning that the structure of the absorbing phase
was considered so far as a relevant parameter.
For example, it was expected that models with infinitely
many absorbing states belong to a different universality class
than DP.
The observed (steady state and dynamical) DP-scaling 
behavior of the pair contact process at criticality 
reveals that the number of absorbing states does not
contribute to the asymptotic scaling behavior at criticality.
For the sake of completeness, we mention that 
different universality classes than DP occur if the 
rapidity reversal is broken, 
e.g.~by quenched 
disorder~\cite{HINRICHSEN_1,JANSSEN_11,JENSEN_10,MOREIRA_1,CAFIERO_1,HOOYBERGHS_1,VOJTA_1},
or additional symmetries~\cite{JANSSEN_1,GRASSBERGER_2} such as 
particle-hole symmetry 
(compact directed percolation~\cite{DOMANY_1,ESSAM_1}),
or particle conservation 
(Manna universality class~\cite{MANNA_2,ROSSI_1,LUEB_26})
or parity conservation (for example branching annihilating
random walks with an even number of offsprings~\cite{ZHONG_1,CARDY_2}).


We would like to thank P.~Grassberger for communicating his results
prior to publication.
Furthermore, we thank H.-K.~Janssen and R.\,M.~Ziff 
for instructive discussions and useful comments on the manuscript.

\newpage


\begin{thebibliography}{10}
\bibitem{LUEB_32_BEM}
\protect{An instructive example is known from equilibrium. Whereas the critical
  exponents of the three-dimensional Ising, XY, and Heisenberg universality
  class differ slightly certain amplitude combinations differ by a factor 1.5
  or even by more than a factor 2 (see \protect\cite{PRIVMAN_2} for an overview). 
  Note that
  amplitude combinations are just particular values of the corresponding
  scaling functions.}
\end{thebibliography}
\end{document}